\documentclass[%
reprint,
 amsmath,amssymb,
 aps,
]{revtex4-2}
\usepackage[bbgreekl]{mathbbol}
\usepackage{graphicx}
\graphicspath{{./figures/}} 
\usepackage{dcolumn}
\usepackage{bm}
\usepackage{slashed}
\usepackage{xcolor}

\usepackage{etoolbox}

\newcommand{\avg}[1]{\left\langle #1 \right\rangle}

\newcommand{\ket}[1]{\left| #1 \right\rangle}
\newcommand{\bra}[1]{\left\langle #1 \right|}
\newcommand{\lrp}[1]{\left( #1 \right)}
\newcommand{\lrb}[1]{\left[ #1 \right]}
\newcommand{\lrc}[1]{\left\{ #1 \right\}}

\newcommand{\bv}[1]{\breve{#1}}

\newcommand{\nonumpar}{\nonumber\\}

\newcommand{\rmT}{\mathrm{T}}

\newcommand{\tr}{\mathrm{Tr}}

\newcommand{\rmEven}{\mathrm{Even}}
\newcommand{\rmOdd}{\mathrm{Odd}}
\newcommand{\rmconst}{\mathrm{const}}

\newcommand{\half}{\frac{1}{2}}
\newcommand{\hfs}{\frac{1}{\sqrt{2}}}

\newcommand{\og}{\omega}
\newcommand{\ld}{\lambda}
\newcommand{\Ld}{\Lambda}
\newcommand{\alp}{\alpha}

\newcommand{\gam}{\gamma}

\newcommand{\barn}{\bar{n}}

\newcommand{\barpsi}{\bar{\psi}}
\newcommand{\barPsi}{\bar{\Psi}}
\newcommand{\barpsip}{\bar{\psi}^{+}}
\newcommand{\barpsim}{\bar{\psi}^{-}}
\newcommand{\psip}{\psi^{+}}
\newcommand{\psim}{\psi^{-}}
\newcommand{\barphi}{\bar{\phi}}
\newcommand{\barPhi}{\bar{\Phi}}

\newcommand{\coc}{\hat{c}^{\dagger}}
\newcommand{\aoc}{\hat{c}}
\newcommand{\cod}{\hat{d}^{\dagger}}
\newcommand{\aod}{\hat{d}}

\newcommand{\coM}{M^\dagger}

\newcommand{\hatU}{\hat{U}}
\newcommand{\cohU}{\hat{U}^\dagger}
\newcommand{\hatH}{\hat{H}}

\newcommand{\hatN}{\hat{N}}

\newcommand{\tldg}{\tilde{g}}
\newcommand{\tldn}{\tilde{n}}

\newcommand{\bvI}{\bv{I}}
\newcommand{\bvG}{\bv{G}}
\newcommand{\bvQ}{\breve{Q}}

\newcommand{\bvPsi}{\bv{\Psi}}

\newcommand{\bvSig}{\bv{\Sigma}}

\newcommand{\bvbPsi}{\bv{\bar{\Psi}}}

\newcommand{\bfI}{{\bf I}}

\newcommand{\bfK}{{\bf K}}

\newcommand{\mcC}{{\mathcal{C}}}
\newcommand{\mcD}{{\mathcal{D}}}

\newcommand{\mcF}{{\mathcal{F}}}

\newcommand{\mcT}{{\mathcal{T}}}
\newcommand{\mcU}{{\mathcal{U}}}

\newcommand{\bbA}{{\mathbb{A}}}
\newcommand{\bbB}{{\mathbb{B}}}
\newcommand{\bbC}{{\mathbb{C}}}
\newcommand{\bbD}{{\mathbb{D}}}

\begin{document}

\preprint{APS/123-QED}

\title{Full counting statistics of the particle currents through a Kitaev chain and the exchange fluctuation theorem}

\author{Fan Zhang}
\affiliation{School of Physics, Peking University, Beijing, 100871, China}

\author{H. T. Quan}\thanks{Corresponding author: htquan@pku.edu.cn}
\affiliation{School of Physics, Peking University, Beijing, 100871, China}
\affiliation{Collaborative Innovation Center of Quantum Matter, Beijing 100871, China}
\affiliation{Frontiers Science Center for Nano-optoelectronics, Peking University, Beijing, 100871, China}



\date{\today}

\begin{abstract}
Exchange fluctuation theorems~(XFTs) describe a fundamental symmetry relation for particle and energy exchange between several systems. Here we study the XFTs of a Kitaev chain connected to two reservoirs at the same temperature but different bias. By varying the parameters in the Kitaev chain model, we calculate analytically the full counting statistics of the transport current and formulate the corresponding XFTs for multiple current components. We also demonstrate the XFTs with numerical results. {\color{black} We find that due to the presence of the $\mcU(1)$ symmetry breaking terms in the Hamiltonian of the Kitaev chain, various forms of the XFTs emerge, and they can be interpreted in terms of various well-known transport processes.}
\end{abstract}


\maketitle
\section{Introduction}\label{sec1Intro}

The fluctuation theorems~(FTs) represent a major advancement in non-equilibrium statistic mechanics, which extend the second law of thermodynamics from inequalities to equalities. The typical form of a FT asserts that there exists an exact relation between the positive entropy production and the negative entropy production, such as 
\begin{align}
	\frac{P(+\Delta\Omega)}{P(-\Delta\Omega)} = e^{\Delta\Omega}, \label{entropy}
\end{align}
where $P(\Delta\Omega)$ denotes the probability distribution of the entropy production. Eq.~(\ref{entropy}) characterizes how much the micro-reversibility is broken and sheds light on the arrow-of-time problem. In addition, the second law $\avg{\Delta\Omega}\geq 0$ is a corollary of the FT (\ref{entropy}).

In the past twenty-five year or so, many FTs have been discovered, for both classical and quantum, as well as both closed and open  systems~\cite{Esposito2009rmp, Campisi2011rmp, Seifert2012rpp, Jarzynski2011an, Zhang2012pr,ge2012pr}.  
In general, FTs can be divided into two categories~\cite{Seifert2012rpp}: the first one concerns the non-equilibrium work, in which the detailed balance condition is satisfied. Examples include Jarzynski's equality~\cite{Jarzynski1997prl} and Crook's FT~\cite{Crooks1999pre}. Parallel to the development of non-equilibrium work FTs, which now extend their domain to arbitrary initial states~\cite{Jarzynski2000jsp,Maragakis2008jpcb} and try to include coherence~\cite{Holmes2019quantum}, the second category of FTs concerning the entropy production~\cite{Evans1993prl, evans1994pre, Gallavotti1995jsp, Gallavotti1995prl, Kurchan1998jpa, Lebowitz1999jsp, Maes1999jsp, Seifert2005prl, Seifert2012rpp}, has also attracted much attention. Among the second category of FTs, a lot of them are relevant to the processes of particle and energy exchange between several systems~\cite{Jarzynski2004prl,Pilgram2004prb, Kindermann2004prb, Saito2008prb,Andrieux2009njp, Utsumi2009prb,Heikkil2009prl, Altland2010prl, Altand2010prb, Sanchez2010prl, Utsumi2010prb, Golubev2011prb, Krause2011prb, Cuetara2011prb, Ganeshan2011prb, Kung2012prx, Gaspard2013book}.
 For example, Jarzynski and W\'{o}jcik discovered a symmetry relation between two bodies initially prepared at different temperatures and they coined the name \textit{exchange fluctuation theorem}~(XFT)~\cite{Jarzynski2004prl}. Later, Andrieux and Gaspard et.~al considered an open quantum system, in which energy and particle currents flow due to the temperature and chemical potential difference. They proved that a transient current FT can be recasted into a steady-state current FT in the long-time limit, in which only the thermodynamic forces or affinities are involved~\cite{Andrieux2009njp}. In Refs.~\cite{Saito2008prb, Utsumi2009prb, Utsumi2014prb}, Utsumi et.~al. found that for electron transport through multiterminal interacting quantum dots under a finite magnetic field, the FT is equivalent to a symmetry in full counting statistics, which not only reproduces the Onsager-Casimir relation, but also suggests universal relations among nonlinear transport coefficients. Thereafter XFTs have been explored in many classical and quantum models, such as the open quantum harmonic chain~\cite{Saito2007prl, Agarwalla2014prb}, coupled quantum spin chains ~\cite{Landi2016pre}, the 2D Ising model~\cite{Piscitelli2008jpa,Piscitelli2009jsm}, a quenched ferromagnet~\cite{Corberi2013jpa}, the spin-boson model~\cite{Nicolin2011jcp}, integrable~\cite{Jason2020scip} and quasi-integrable systems~\cite{Goldfriend2018epl}. Most of these works arrive at a Gallavati-Cohen-type steady-state FT~\cite{Gallavotti1995prl} or a transient FT~\cite{Jarzynski2004prl}.  But in some cases, the XFTs need modifications. For example, when the environment temperature is under periodic control~\cite{Watanabe2017pre}, or when correlations are added to the initial state~\cite{Jeon2016pre}, extended forms of XFT were found. Recently, a tightest and saturable matrix-valued thermodynamic uncertainty relation is derived from the XFTs, which imposes strict restrictions on the fluctuations of thermodynamic currents \cite{Barato2015prl,Timpanaro2019prl}.
Experimentally, the XFTs have been tested in various systems, such as Brownian particles~\cite{Gomez2011prl,Berut2016prl}, classical conductors~\cite{Ciliberto2013prl}, quantum dot system~\cite{Utsumi2009prb,Kung2012prx}, the NV center spin qubit~\cite{Hernandez2020prr} and NMR setups~\cite{Pal2019pra, Pal2020prr}.

\begin{figure}
\includegraphics[scale=0.7]{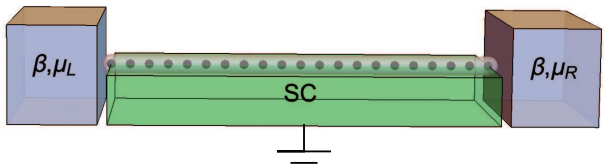}
\caption{The setup. A nanowire is put above a s-wave superconductor~(SC, green) and connected to two reservoirs~(blue). The SC is grounded while the chemical potentials of the two reservoirs are $\mu_L$ and $\mu_R$, respectively. The whole system is in a uniform temperature $T=1/\beta$.}
\label{setup}
\end{figure}

 Despite of these efforts, {\color{black}to the best of our knowledge, the particle number conservation is always guaranteed in the previous studies. However, in some low-energy effective Hamiltonians, the particle number conversation, associated with the global $\mcU(1)$ symmetry, can be explicitly broken, such as the BCS Hamiltonian. It would be interesting to explore the influence of the $\mcU(1)$ symmetry breaking on the distribution of the particle currents as well as the form of the XFTs.} In this article, we explore these problems by studying a 1D open Kitaev chain in the framework of full counting statistics~(FCS). By using Keldysh formalism, we analytically calculate the characteristic function of the charge currents. In analogy to the partition function, which encodes essential information about equilibrium state, the characteristic function of the {\color{black}particle} currents encodes essential information about the nonequilibrium process. From our results, one can see that the multiple {\color{black}particle} current components in the Kitaev chain lead to various {\color{black}forms}  of XFTs, {\color{black} as a result of the $\mcU(1)$ symmetry breaking.}
 
This article is organized as follows. In Sec.~II, we introduce the 1D Kitaev chain model and we use Keldysh technique to calculate the characteristic function. In Sec. III, we discuss our analytical results of a short Kitaev chain and show the well-known Andreev reflections giving rise to  non-conventional XFTs. Conclusions are drawn in Sec. IV.

\section{1D Kitaev Model}\label{sec2model}
The 1D Kitaev chain is a quadratic model, in which Majorana zero modes~(MZMs) exist when the system is in the topological superconductor phase~\cite{kitaev2001ps}. It can be realized by putting a semiconductor nanowire on a s-wave superconductor~\cite{Sau2010prl,Lutchyn2010prl,Mourik2012science, Lutchyn2018nrev}. Due to the proximity effect, Cooper pairs can tunnel into the nanowire. The low-energy effective Hamiltonian of the nanowire can be described by the Kitaev model. In condensed matter physics, topological properties of the system have attracted a lot of attention in recent years, since MZMs may serve as a platform to realize the fault-tolerant topological quantum computation. It's worth mentioning that braiding the MZMs to realize the logic gate is still a big challenge in this field~\cite{sarma2015npj}. Here, we restrict ourselves to the discussion of the distribution function of the currents and the fluctuation theorems. Nevertheless, we won't touch MZMs since it's not the focus of our current study.

A schematic picture of the setup is given in Fig.~\ref{setup}. The chain is connected to the left and the right reservoirs by the first and the last lattice site respectively. The whole system is in a uniform temperature $T$, but the chemical potential of the two reserviors are different. They are denoted by $\mu_L$ and $\mu_R$ respectively. The total Hamiltonian of the system consists of five parts: the Hamiltonian of the left and the right reserviors $H_L$, $H_R$; the Hamiltonian of the Kitaev chain $H_K$; and the interaction Hamiltonians $H_{LK}$, $H_{KR}$.
\begin{align}
	\hatH = \hatH_L + \hatH_K + \hatH_R + \hatH_{LK} + \hatH_{KR}
\end{align}
where the Hamiltonian for every part is given by
\begin{align}
\hatH_K &= -\mu \sum_{j=1}^N \lrp{\coc_j \aoc_j-\half} \nonumpar
	&~~~+ \sum_{j=1}^{N-1}(-\eta~\coc_{j}\aoc_{j+1} +\Delta ~ \aoc_{j}\aoc_{j+1} + h.c),
			\nonumber \\
\hatH_L	&= \sum_j \lrp{\hbar \og_{Lj}-\mu_L }\coc_{Lj} \aoc_{Lj}, \nonumpar
\hatH_R	&= \sum_j \lrp{\hbar \og_{Rj}-\mu_R }\coc_{Rj} \aoc_{Rj}, \nonumpar 
\hatH_{LK} &=\sum_j \ld_{Lj} \lrp{\coc_{Lj} \aoc_1 + \coc_1 \aoc_{Lj}},\nonumpar
\hatH_{KR} &= \sum_j \ld_{Rj} \lrp{\coc_{Rj} \aoc_N + \coc_N \aoc_{Rj}}.
\end{align}
Here $\mu$ is the chemical potential, $\eta$ is the hopping amplitude and $\Delta$ is the superconducting order parameter. $\aoc_n, \aoc_{Lj}$ and $\aoc_{Rj}$ are the annihilation operators for an electron on the $n$-th site, the left and the right reservoirs respectively. $\lambda_{Lj}~(\lambda_{Rj})$ characterizes the tunneling strength between the left~(right) reservoir and the first~(last) site of the chain. $\hbar\omega_{Lj}$~($\hbar\omega_{Rj}$) denotes the energy of the $j$-th state of the left~(right) reservoir. From now on, we adapt natural units~($\hbar = e = k_B=1$), so a conductance of $1 / \pi$ represents $2 e^2 / h$ in SI units.


We are interested in the charge transport in a period of time $\lrb{0, \tau}$. In the two-point measurement~(TPM) scheme, two projective measurements over the particle number operator $\hat{N}_L = \sum_j \coc_{Lj}\aoc_{Lj}$ of the  left reservoir  are applied at the initial time $t=0$ and the final time $t=\tau$. The initial state is assumed to be a product state $\rho_0 = \rho_L^{eq}\otimes \rho_K^{eq}\otimes\rho_R^{eq}$. The probability distribution of the charge transported from the left reservoir during time $\lrb{0, \tau}$ is given by
{\color{black} 
\begin{align}
	P(q) = \sum_{n_i,n_f}\sum_{\sigma,\sigma'}&\delta\lrb{q-\lrp{n_f-n_i}}p(n_{f},\sigma'|n_{i},\sigma) \times \nonumpar
	 & \bra{n_{i},\sigma}\rho_0\ket{n_{i},\sigma},
\end{align}
where $n_{i}$ and $n_{f}$ are the initial and the finial particle numbers in the left reservoir and $\sigma, \sigma'$ label the many-body degeneracies for a fixed particle number. $p(n_{f},\sigma'|n_{i},\sigma)=\bra{n_{f},\sigma'}\tr_{K,R}\lrb{\hatU(\tau,0)\hat{P}(n_i, \sigma)\otimes\rho_K^{eq}\otimes\rho_R^{eq}\hatU^\dagger(\tau,0)}\ket{n_{f},\sigma'}$ denotes the transition probability, in which $\hat{P}(n_i, \sigma)=\ket{n_i,\sigma}\bra{n_i,\sigma}$ is the projective operator of the eigenstate $\ket{n_i, \sigma}$.} $\hatU(\tau,0) = \mcT\exp\lrp{-i\int_{0}^\tau \hatH(t)dt}$ is the unitary evolution operator where $\mcT$ is the time-ordering operator.
It's more convenient to consider the characteristic function~(CF), namely, the Fourier transform of the probability distribution which can be expressed in a trace form,
\begin{align}
	Z(\xi) &= \int dq~e^{i\xi q}P(q) \nonumpar
	&= \tr{\lrb{\cohU(\tau,0) e^{i\xi \hatN_L}\hatU(\tau,0)e^{-i\xi \hatN_L}\rho_0}}.\label{cfq}
\end{align}
Since the initial state is a product state which commutes with $\hatN_L$, we can further rewrite the CF as
\begin{align}
	Z(\xi) &= \tr\lrb{e^{-i\xi\hatN_L/2}\cohU(\tau,0)e^{i\xi\hatN_L}\hatU(\tau,0)e^{-i\xi\hatN_L/2}\rho_0} \nonumpar
	&= \tr\lrb{\cohU_{-\xi/2}(\tau,0)\hatU_{\xi/2}(\tau,0)\rho_0},
\end{align}
where the modified evolution operator is given by 
\begin{align}
	\hatU_x(t,t') &= e^{ix\hatN_L}\hatU(t,t')e^{-ix\hatN_L} \nonumpar
	&= \mcT \exp\lrc{-i\int_{t'}^t \hatH_x(t') dt'}
\end{align}
with the modified Hamiltonian  $\hatH_x(t)$ 
\begin{align}
	\hatH_x(t) &= e^{ix\hatN_L}\hatH(t)e^{-ix\hatN_L} \\
		&= \hatH + \sum_{j}\lambda_{Lj} \lrb{\lrp{e^{ix}-1}\coc_{Lj}\aoc_1 +\lrp{e^{-ix}-1}\coc_1\aoc_{Lj}}. \nonumber
\end{align}
Please note that the CF is also called the generating function in the field of quantum transport since it generates the irreducible cumulants by taking the derivative of $\xi$~\cite{Gaspard2013book, Kamenev2011book},
\begin{align}
	\avg{\avg{q^n}} = (-i)^n\frac{\partial^n \ln Z(\xi)}{\partial \xi^n}\bigg|_{\xi=0}.
\end{align}
We also note that when $\xi\ll 1$, the modified Hamiltonian can be approximated as 
\begin{align}
	\hatH_x(t)&\approx \hatH + \sum_{j}ix\lambda_{Lj} \lrp{\coc_{Lj}\aoc_1 -\coc_1\aoc_{Lj}}\nonumpar
	& = \hatH - x\hat{I}_L,  \label{nazarov}
\end{align}
where the current operator $\hat{I}$ is defined as
\begin{align}
	\hat{I}_L &\equiv \frac{d\hatN_L}{dt} = i\lrb{\hatH_L, \hatN_L} = \sum_{j}i\lambda_{Lj} \lrb{\coc_1\aoc_{Lj}-\coc_{Lj}\aoc_1 }.\nonumber
\end{align}
Eq.~(\ref{nazarov}) is in the same form as defined by Nazarov~\cite{Nazarov2003epj}. However, due to the approximation~(cut-off) in Eq.~(\ref{nazarov}), this form doesn't possess Gallavotti-Cohen symmetry and gives only the first two cumulants correctly~\cite{Agarwalla2013thesis}.

 Various methods have been developed to calculate the CF, such as the scattering matrix approach~\cite{levitov1993JETP}, the stochastic path integral in semi-classical regime~\cite{Pilgram2003prl}, the quantum master equation approach~\cite{Ren2010prl} and the Keldysh Green's function approach~\cite{Nazarov2003epj}. Due to the quadratic form of the Hamiltonian of the Kitaev chain, we apply the Keldysh technique which can simplify the calculation by utilizing the Gaussian integral. In the Keldysh formulation, the CF can be written as a functional integral along the Schwinger-Keldysh contour~\cite{Kamenev2011book},
 \begin{align}
	Z(\xi) &= \tr\lrb{\rho_0 \exp\lrp{-i\int_\mcC \hatH_{\xi(t)/2}(t)dt}}\nonumpar
	&= \int \mcD\lrb{\barpsi,\psi,\barphi,\phi}e^{iS\lrb{\barpsi,\psi,\barphi,\phi, \xi}}
\end{align}
where the contour $\mcC$ consists of a forward path $(0,\tau)$ and a backward path $(\tau,0)$. And the counting field $\xi(t)$ is set to be $\xi(t)=\pm\xi$~(plus for the forward and minus for the backward). In the last step, we insert fermionic coherent state $\psi,\phi,\barpsi,\barphi$ and the action consists of three parts, namely, 
\begin{align}
	S &= S_K + S_{\mathrm{res}} + S_{\rmT}, \nonumpar
	S_K &= \int_{\mcC} dt ~\barPsi(t) \lrp{i\frac{d}{dt}  - \bfK}\Psi (t),  \nonumpar
	S_{\mathrm{res}} &= \sum_{j}\sum_{\alpha=L,R} \int_{\mcC}dt ~\bar{\phi}_{\alpha j} \lrp{i\frac{d}{dt} - \og_{\alpha j}}\phi_{\alpha j}, 
	\nonumpar
	S_{\rmT} 
	&= -\sum_j \int_\mcC dt ~\ld_{Lj}\lrb{\barPhi_{Lj}(t)M(t)\Psi_1(t) + \barPsi_1(t)\coM(t)\Phi_{Lj}(t)} \nonumpar
	&\quad -\sum_j \int_\mcC dt ~\ld_{Rj}\lrb{\barPhi_{Rj}(t)\Psi_N(t) + \barPsi_N(t)\Phi_{Rj}(t)}, \label{totaction}
\end{align}
where $\Psi^T=\lrp{\Psi_1,\ldots, \Psi_N}^T$, $\Psi_i^T = \hfs \lrp{\psi_i,\barpsi_i}^T$, $\Phi_{\alpha j}^T= \hfs\lrp{\phi_{\alpha j},\barphi_{\alpha j}}^T$ are Grassmann numbers in Nambu space for the chain and the reservoirs respectively. {\color{black}$\bfK$ in Eq.~(\ref{totaction}) is the representation of the Hamiltonian of the Kitaev chain in Nambu space, defined as a $2N\times 2N$ matrix
\begin{align}
	\bfK = -\begin{pmatrix}
		\mu\sigma_3 & D  & 0 & \ldots & \ldots & 0 \\
		D^T & \mu\sigma_3 & D & 0 & \ldots & 0 \\
		0 & D^T & \ddots & \ddots & \ddots & \vdots \\
		\vdots & \ddots & \ddots & \ddots & D & 0 \\
		0 & \ldots & 0 & D^T & \mu\sigma_3 & D \\
		0 & \ldots & \ldots & 0 & D^T & \mu\sigma_3
	\end{pmatrix}
\end{align} 
in terms of the Pauli matrix $\sigma_3$ and the $2 \times 2$ matrix 
\begin{align}
	D \equiv \eta\sigma_3+i\Delta\sigma_2 = \begin{pmatrix}
		\eta & \Delta \\
		-\Delta & -\eta
	\end{pmatrix}.
\end{align}}
{\color{black}$M(t)$ in Eq.~(\ref{totaction}) is the counting field matrix} 
\begin{align}
	M(t) = \begin{pmatrix}
		e^{i\xi(t)/2} & 0 \\
		0 & -e^{-i\xi(t)/2}
	\end{pmatrix}.
\end{align}
We project the contour time to the real-time axis which doubles the Grassmann number $\psi_i\to \bv{\psi}_i\equiv (\psip_i, \psim_i)$. Here the breve indicates the Grassmann number or operators in the Keldysh space and the superscript plus~(minus) represents the Grassmann numbers are defined in the forward~(backward) path. Then we perform the Larkin-Ovchinnikov~(L-O) rotation in the Keldysh space, i.e., 
\begin{align}
	\bv{\Psi}'_i &= \begin{pmatrix}
	\Ld_1 & 0\\
	0 & \Ld_1
\end{pmatrix}\begin{pmatrix}
	\psip_i \\ \psim_i \\ \barpsip_i \\ \barpsim_i
\end{pmatrix} \equiv \begin{pmatrix}
	\psi_i^1 \\ \psi_i^2 \\ \barpsi_i^2 \\ \barpsi_i^1
\end{pmatrix}, \nonumpar
\bv{\barPsi}'_i &= \begin{pmatrix} \barpsip_i & \barpsim_i & \psip_i & \psim_i \end{pmatrix}\begin{pmatrix}
	\Ld_2 & 0\\
	0 & \Ld_2
\end{pmatrix}
\equiv \begin{pmatrix} \barpsi_i^1 & \barpsi_i^2 & \psi_i^2 & \psi_i^1 \end{pmatrix}
\end{align}
where the rotation matrices are defined as 
\begin{align}
	\Ld_1 = \hfs\begin{pmatrix} 1 & 1 \\ 1 & -1 \end{pmatrix},\quad 
	\Ld_2 =  \hfs \begin{pmatrix} 1 & 1 \\ -1 & 1 \end{pmatrix}.
\end{align}
The reservoir action becomes
\begin{align}
	S_{\mathrm{res}} &= \sum_{j}\sum_{\alpha=L,R}\int dtdt'~\bv{\barPhi}_{\alpha j}(t)\cdot \bv{Q}_{0,\alpha j}^{-1}(t,t')\bv{\Phi}_{\alpha j}(t') 
\end{align}
with $\bv{\barPhi}_{\alpha j}(t) = \hfs \begin{pmatrix}
	\barphi^1 & \barphi^2 & \phi^2 & \phi^1
		\end{pmatrix}_{\alpha j}$ and the free reservoir Green's function
\begin{align}
	\bv{Q}_{0,\alpha j}(k, \omega) = \begin{pmatrix}
		g^0_{\alpha j}(k, \omega)  & 0 \\
		0 & \tldg^{0}_{\alpha j}(k, \omega)
	\end{pmatrix}.
\end{align}
The particle Green's function $g^0_{\alpha j}(k, \omega)$ and the hole Green's function $\tldg^0_{\alpha j}(k, \omega)$ are given by 
{\color{black}
\begin{align}
	g^0_{\alpha j}(k, \omega) &= \begin{pmatrix}
		\frac{1}{\omega - \omega_{\alp k}' +i\eta} & -2\pi i\lrb{1- 2n(\omega)}\delta(\omega-\omega_{\alp k}') \\
		0 & \frac{1}{\omega - \omega_{\alp k}' -i\eta}
	\end{pmatrix}, \nonumpar
	\tldg^0_{\alpha j}(k, \omega) &= \begin{pmatrix}
		\frac{1}{\omega + \omega_{\alp k}' +i\eta} & -2\pi i\lrb{1- 2\tldn (\omega)}\delta(\omega+\omega_{\alp k}') \\
		0 & \frac{1}{\omega + \omega_{\alp k}' -i\eta}
	\end{pmatrix},\nonumpar
	\tilde{n}(\omega) &= 1-n(-\omega) = \frac{1}{1+e^{\beta\lrp{\omega+\mu}}},
\end{align}
where $\omega_{\alp k}'\equiv \omega_{\alp k}-\mu_\alp$.} Thanks to the quadratic form of the reservoir action, we can integrate out the reservoir degrees of freedom and get the self-energy 
{\color{black}
\begin{align}
\bv\Sigma_L^{\xi}(t,t') &= \sum_j \lambda_{Lj}^2 \bv{M}^\dagger \cdot\bvQ_{0,Lj}\cdot \bv{M}, \nonumpar 
\bv\Sigma_R^0(t, t') &= \sum_j \ld_{Rj}^2 \lrp{\Ld_2\otimes I_2} \cdot \cdot\bvQ_{0,Rj}\cdot \lrp{\Ld_1\otimes I_2},
\end{align}}
where the counting field matrix in the Keldysh space is 
\begin{align}
	\bv{M}^\dagger &= \lrp{\Ld_2\otimes I_2} \cdot \coM \cdot \lrp{\Ld_1\otimes I_2}, \nonumpar
	\bv{M} &= \lrp{\Ld_2\otimes I_2} \cdot M \cdot \lrp{\Ld_1\otimes I_2}
\end{align}
with $I_2$ the $2\times 2$ identity matrix.
The effective action of the Kitaev chain is 
\begin{align}
	S_{\mathrm{eff}}(\bvPsi', \bvbPsi', \xi) =  &\iint dtdt' ~\bvbPsi'(t) \bigg[i\delta(t, t')\lrp{\frac{d}{dt'} - \bfK}\nonumpar
	&-\bv{\Sigma}_L^{\xi}(t, t') - \bv{\Sigma}_R^0(t, t')\bigg]\bvPsi'(t').
\end{align}
If we only consider the fermionic states near the Fermi surface, we can linearize the fermionic dispersion relation $\omega_k\approx v_F k$ where $v_F$ is the Fermi velocity. We can also assume the coupling strength $\lambda_{\alpha j}$~($\alpha=L,R$) are approximately constant $\lambda_\alpha$ for these states~{\color{black}(wide-band approximation)},  which leads to
\begin{align}
	\sum_j \lambda_{\alpha j}^2 g^0_{\alpha j}(k,\omega) \approx -i\Gamma_\alpha \begin{pmatrix}
		1 & 2(1-2n(\omega)) \\
		0 & -1
		\end{pmatrix}, \nonumpar
	\sum_j \lambda_{\alpha j}^2 \tldg^0_{\alpha j}(k,\omega) \approx -i\Gamma_\alpha \begin{pmatrix}
		1 & 2(1-2\tilde{n}(\omega)) \\
		0 & -1
		\end{pmatrix}	
\end{align}
with {\color{black}$\Gamma_\alpha \equiv \lambda_\alpha^2/(v_F)$}. 

In the long-time limit $\tau\to \infty$, we can turn to the frequency domain and obtain the final form of the CF
\begin{align}
	\lim_{\tau\to\infty}Z(\xi) &= \lim_{\tau\to\infty}\prod_\omega \sqrt{\frac{\det\lrb{\bvI -\bvG(\omega)\bv{\Sigma}(\xi,\omega)}}{\det\lrb{\bvI -\bvG(\omega){\color{black}\bv{\Sigma}(0,\omega)}}}} \nonumpar
	&= \lim_{\tau\to\infty}\prod_\omega \sqrt{Z(\xi,\omega)}  \label{eqGF}
\end{align} 
{\color{black}with Green's function and self-energy 
\begin{align}
	\bvG(\omega) = (\omega \bfI - \bfK)^{-1},\quad \bvSig(\xi, \omega) = \bv{\Sigma}_L^{\xi}(\omega) + \bv{\Sigma}_R(\omega). 
\end{align}}
A more frequently used quantity in the long-time limit is the cumulant generating function~(CGF), defined by
{\color{black}
\begin{align}
	\mcF(\xi) &= \lim_{\tau\to\infty}\frac{1}{\tau}\ln Z(\xi) \nonumpar
	&=\half \int_{-\infty}^{\infty}\frac{d\omega}{2\pi}\ln \det\lrb{\bvI -\bvG(\omega)\bv{\Sigma}(\xi,\omega)}+\rmconst, \label{eqCGF}
\end{align} 
where we have absorbed the $\xi-$independent term~(the denominator in Eq.~(\ref{eqGF})) into the $\rmconst$, because it has no effect on the cumulants and the FTs.} The existence of CGF implies that the current satisfies a large deviation principle, i.e., probability $P(q)\sim \exp\lrb{-\tau h(q/\tau)}$ for large $\tau$ with the rate function $h(q/\tau) = \mcF(\xi^*)-i\xi^* q/\tau$, where $\xi^*$ is the solution of the saddle-point equation $d\mcF(\xi^*)/d\xi^* = iq/\tau$~{\color{black}\cite{Hugo2009PR}. As for the numerical calculation, it's more convenient to adapt the inverse Fourier transform to obtain $P(q)$ and $h(q/\tau)$.} These two relations Eq.~(\ref{eqGF}) and Eq.~(\ref{eqCGF}) are valid for arbitrary number of sites in the Kitaev chain. In the following, we will only consider the CF in the long-time limit $\tau\to\infty$. 
\begin{figure}
\includegraphics[scale=0.7]{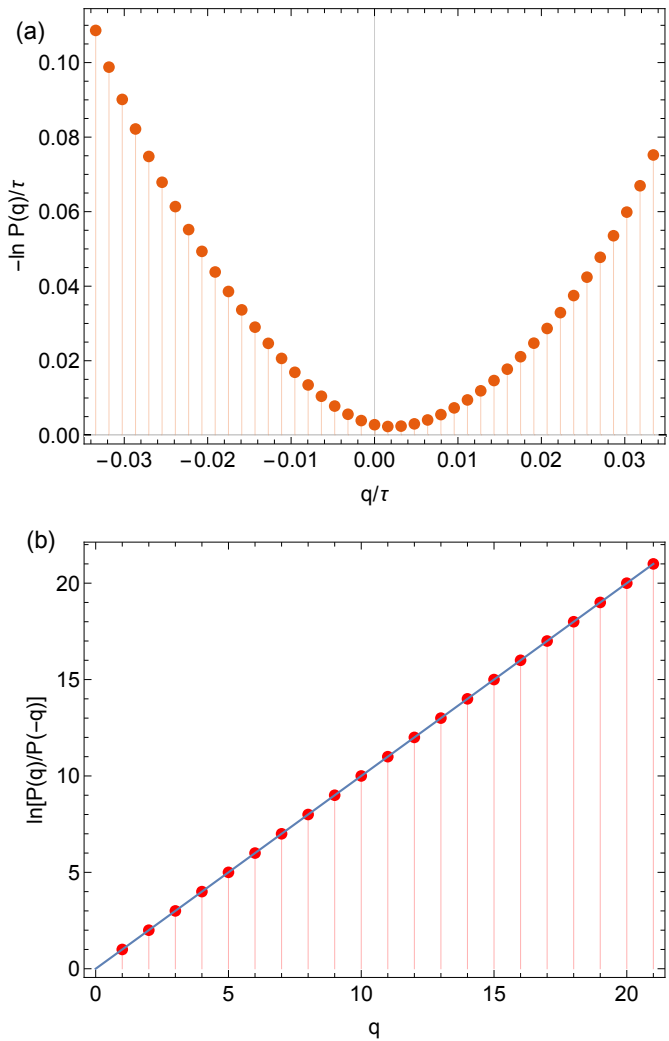}
\caption{The rate function and the XFT for 10 sites in the trivial case. The parameters are chosen to be : $\mu=\eta=1, \Delta=0$, $\Gamma_L=\Gamma_R=0.3$, $\mu_L-\mu_R=0.1$, $\beta=10$, $\Delta\omega \equiv 2\pi/\tau=0.01$. (a) and (b) show $-\ln P(q)/\tau$ and $\ln P(q)/\ln P(-q)$ as functions of $q$ respectively. The solid line is $\beta(\mu_L-\mu_R)q = q$ and it is a manifestation of the XFT~(\ref{XFTtr}).}
\label{norm10u1d0}
\end{figure}

\section{Exchange FT in the open Kitaev chain}
In this section, we will formulate the FTs and calculate the CF  analytically. We would like to emphasis that the validity of the FTs is independent of the site number in the Kitaev chain, but the expression of the CF depends on the site number. As a result, the cumulants, such as the average current and the shot noise explicitly depend on the site number. For simplicity, we consider a short chain of, for example, only $3$ sites. In principle, generalization to more sites is straightforward. It's worth mentioning that the expression of the CF for a large site number can be obtained by the same method as that for a small site number. But in practice, the calculation is extremely tedious for a large site number. We would like to point out that in the study of quantum transport, the system is usually assumed in the thermodynamic limit; and the continuum limit can be used to simplify the calculation of the CF. But for a system in the thermodynamic limit, the fluctuations are suppressed and {\color{black}can't reach a steady state. For these reasons, the thermodynamic limit is not suitable for the study of FTs.} 
\begin{figure*}
\includegraphics[scale=0.6]{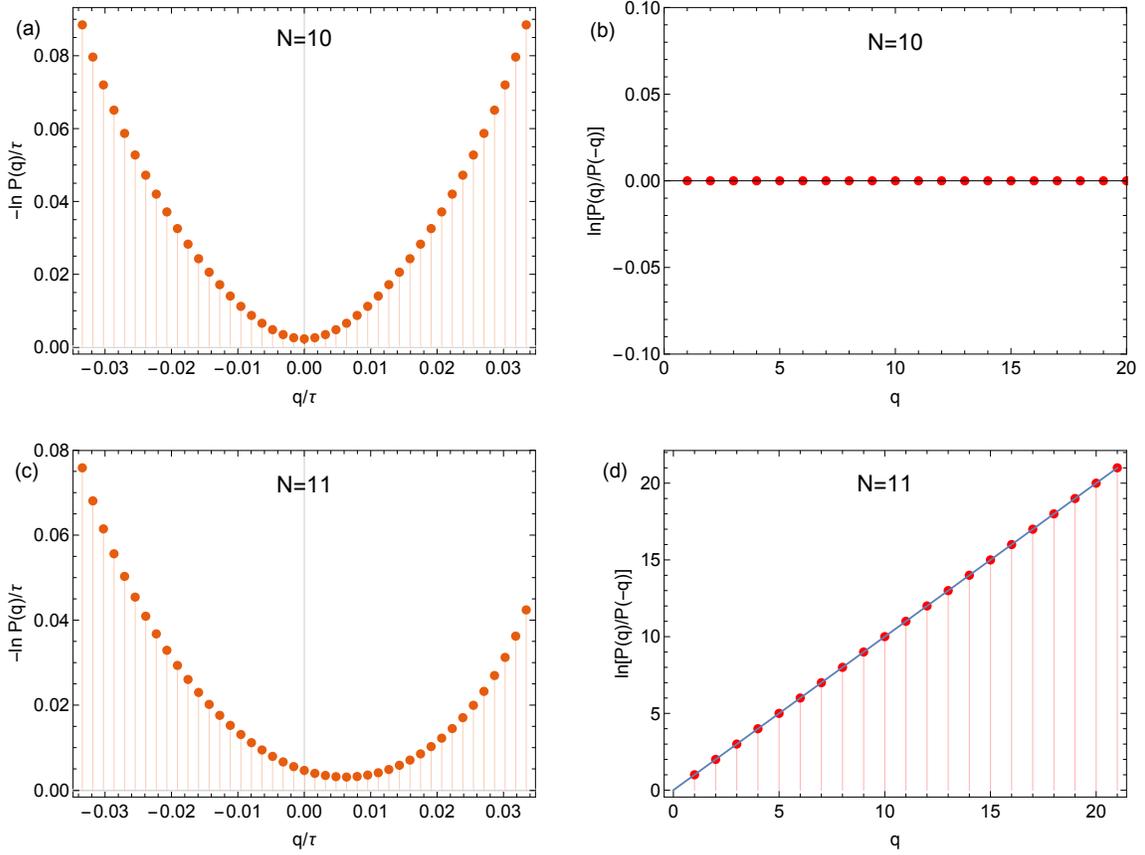}
\caption{The rate functions and the XFTs for even/odd parity systems. The parameters are: $\Delta=1$, $\mu=\eta=0$, $\Gamma_L=\Gamma_R=0.3$, $\mu_L=-\mu_R=0.05$, $\beta=10$, $\Delta\omega \equiv 2\pi/\tau=0.01$. (a), (c): The rate function $-\ln P(q)/\tau$ for $10, 11$ sites, respectively. (b), (d): The ratio $\ln P(q)/\ln P(-q)$ for $10, 11$ sites, respectively. The solid line in (d) represents $\beta (\mu_L-\mu_R)q=q$.}
\label{sc1011u0d1T0}
\end{figure*}
In the following, we will consider four cases: i). trivial case when the chain has no pairing potential; ii). pairing case when the hopping amplitude is zero and the current is carried by Cooper pairs; iii). Majorana case when $\mu=0,\Delta=\eta$, i.e., the system possesses Majorana modes; iv). general case.
\subsection{Trivial case}
We take the pairing potential $\Delta=0$, then the chain is a trivial conductor chain. Similar model can be found at in Ref.~\cite{Esposito2009rmp}~(see Eq.~(126) therein). The CF for our 3-site model at a finite temperature is 
{\color{black}
\begin{align}
	Z_3(\xi,\omega) 
	= &\lrb{\bbA_{1-}(\omega)+\bbC_{1-}(\omega)\lrp{\barn_{1e} n_{2e} e^{-i\xi} +n_{1e}\barn_{2e} e^{i\xi}}}\times \nonumpar
	&\lrb{\bbA_{1+}(\omega)+\bbC_{1+}(\omega)\lrp{n_{1h}\barn_{2h}e^{-i\xi}+ \barn_{1h}n_{2h}e^{i\xi}}}, \label{trcf}
\end{align}
}
which is the Levitov-Lesovik formula for each orbits~\cite{Levitov1996jmp}. 
The coefficients $\bbA_{1\pm}(\omega), \bbC_{1\pm}(\omega)$ are functions of parameters of the system such as $\mu, \eta,\Gamma_\alpha$ but are $\xi$-independent~(see Appendix~\ref{coefficient}). $n_{1e}$ and $n_{1h}$~($n_{2e}$ and $n_{2h}$) are the electron and the hole occupation number in the left~(right) reservoir while $\barn(\omega)\equiv 1-n(\omega)$,
\begin{align}	
	n_{1e}(\omega) = \frac{1}{e^{\beta(\omega-\mu_L)}+1}, \quad  \barn_{1e}(\omega) \equiv 1-n_{1e}(\omega), \nonumpar
	n_{1h}(\omega) = \frac{1}{e^{\beta(\omega+\mu_L)}+1}, \quad  \barn_{1h}(\omega) \equiv 1-n_{1h}(\omega).
\end{align}
Obviously, since 
\begin{align}
\barn_{1e}n_{2e} = e^{-\beta(\mu_L-\mu_R)}n_{1e}\barn_{2e},
\end{align}  
the CF satisfies a symmetry
\begin{align}
	Z_3(\xi) &= Z_3(-\xi + i\beta(\mu_L-\mu_R)).
\end{align}
As a result, the CGF~(\ref{eqCGF}) satisfies the symmetry
\begin{align}
	\mcF_3(\xi) &= \mcF_3(-\xi + i\beta (\mu_L-\mu_R)).
\end{align}
This symmetry immediately implies the well-known steady-state XFT~(Eq.~(141) in Ref.~\cite{Esposito2009rmp} and Eq.~(85) in Ref.~\cite{Campisi2011rmp})
{\color{black}
\begin{align}
	\lim_{\tau\to\infty}\frac{P(q)}{P(-q)} &= e^{q\beta (\mu_L-\mu_R)}.\label{XFTtr}
\end{align}}

The average current can be obtained by taking derivative of the CGF and we reproduce the Landauer-Buttiker expression~\cite{Blanter2000PR}
\begin{align}
	\avg{I} &= -i\frac{\partial}{\partial \xi}\mcF(\xi)\bigg|_{\xi=0} \nonumpar
	&= \int \frac{d\omega}{4\pi}\bigg[\frac{\bbC_{1+}(\omega)\lrp{n_{1e}\barn_{2e}-\barn_{1e}n_{2e}}}{\bbA_{1+}(\omega)+\bbC_{1+}\lrp{\omega}\lrp{\barn_{1e} n_{2e} +n_{1e}\barn_{2e}}} \nonumpar
	&\quad\quad + \frac{\bbC_{1-}\lrp{\omega}\lrp{\barn_{1h}n_{2h}-n_{1h}\barn_{2h}}}{\bbA_{1-}(\omega)+\bbC_{1-}(\omega)\lrp{\barn_{1h} n_{2h} +n_{1h}\barn_{2h}}}\bigg]. \label{triI}
\end{align}
Eq.~(\ref{triI}) shows the fundamental transport process is an electron transferred from the left reservoir to the chain and meanwhile an electron is transferred from the chain to the right reservoir, thus preserving the charge number in the chain. From Eq.~(\ref{eqGF}), we can numerically calculate the probability distribution $P(q)$ for more than $3$ sites.  In Fig.~\ref{norm10u1d0}, we plot the rate function $-\ln P(q)/\tau$ and the ratio $\ln P(q)/\ln P(-q)$ for $10$ sites. 

\begin{figure}
\includegraphics[scale=0.7]{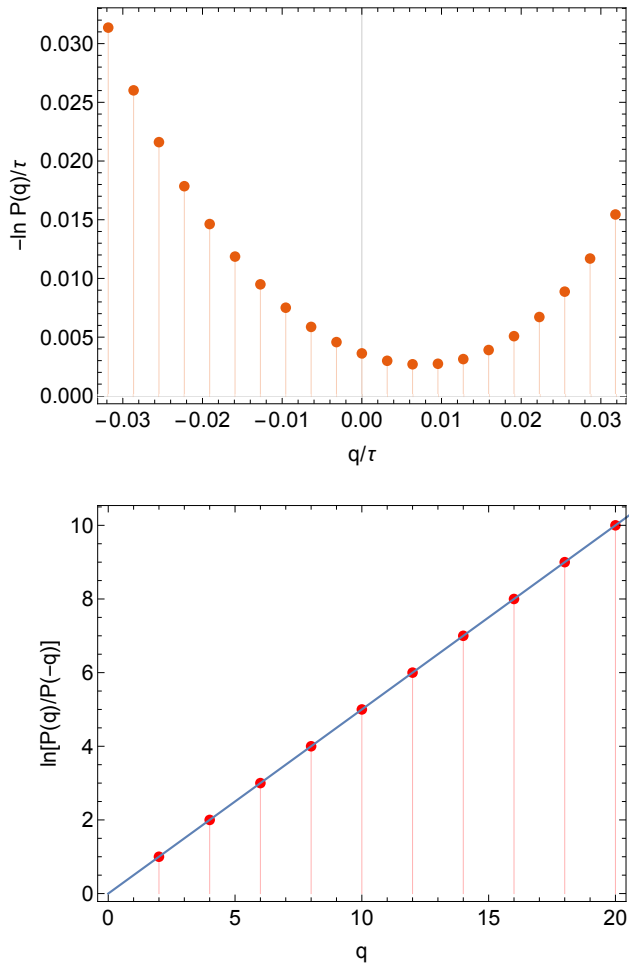}
\caption{The rate function and the XFT for 10 sites in the Majorana case. The parameters are: $\mu=0$, $\Delta=\eta=1$, $\Gamma_L=\Gamma_R=0.3$, $\mu_L=-\mu_R=0.05$, $\beta=10$, $\Delta\omega \equiv 2\pi/\tau=0.01$. (a) The rate function $-\ln P(q)/\tau$. (b) The ratio $\ln P(q)/\ln P(-q)$  shows that only {\color{black}even}-number particles can be transported. The solid line is $\beta \mu_L q=0.5q$.}
\label{tscu0d1}
\end{figure}

\subsection{Pairing case}
We take the hopping amplitude $\eta=0$, and introduce the non-zero paring potential $\Delta$ {\color{black}which explicitly breaks the global $\mcU(1)$ symmetry.} Then the current should be carried only by Cooper pairs. We find that there are subtleties in this case, i.e., the CF exhibits distinct features for odd-number and even-number sites. As examples, we calculate the CF for both $3$ sites and $4$ sites
{\color{black}
\begin{align}
	Z_3(\xi, \omega) = &\lrb{\bbA_{2-}(\omega)+\bbC_{2-}(\omega)\lrp{\barn_{1e} n_{2e} e^{-i\xi} +n_{1e}\barn_{2e} e^{i\xi}}}\times \nonumpar
	&\lrb{\bbA_{2+}(\omega)+\bbC_{2+}(\omega)\lrp{n_{1h}\barn_{2h}e^{-i\xi}+ \barn_{1h}n_{2h}e^{i\xi}}}, \label{pair3cf} \\
	Z_4(\xi, \omega) = &\lrb{\bbA'_{2-}(\omega)+\bbC'_{2-}(\omega)\lrp{\barn_{1e} n_{2h} e^{-i\xi} +n_{1e}\barn_{2h} e^{i\xi}}}\times \nonumpar
	&\lrb{\bbA'_{2+}(\omega)+\bbC'_{2+}(\omega)\lrp{n_{1h}\barn_{2e}e^{-i\xi}+ \barn_{1h}n_{2e}e^{i\xi}}}. \label{pair4cf}
\end{align}}
The $3$-site CF $Z_3(\xi,\omega)$ still describes a normal transport process which is the same as the trivial case except that the coefficients $\bbA_{2\pm}, \bbC_{2\pm}$ are different from $\bbA_{1\pm}, \bbC_{1\pm}$~(see Appendix~\ref{coefficient}). The XFT~(\ref{XFTtr}) holds as usual. However, for the $4$-site case, the CF $Z_4(\xi,\omega)$ is quite different from that of 3-site case. It describes an electron moves from the left reservoir to the chain and a hole moves from the chain to the right reservoir. Meanwhile, a Cooper pair is formed in the chain and is injected into the superconductor. It's called crossed Andreev reflection~(CAR)~\cite{Nilsson2008prl, Law2009prl}. In this case the system can be viewed as a three terminal system: the chain can exchange particles with two reservoirs and the proximate superconductor. The equilibrium relation
{\color{black}
\begin{align}
	\barn_{1e}n_{2h}&= e^{-\beta\lrp{\mu_L+\mu_R}}n_{1e}\barn_{2h}, \nonumpar
	n_{1h}\barn_{2e}&= e^{-\beta\lrp{\mu_L+\mu_R}}\barn_{1h}n_{2e}
\end{align}}
implies the symmetry of the CF 
\begin{align}
	Z_4(\xi,\omega) = Z_4\lrb{-\xi+i\beta\lrp{\mu_L+\mu_R}},
\end{align}
which further implies the symmetry of the CGF and the XFT
\begin{align}
	\mcF_4(\xi) &= \mcF_4\lrb{-\xi+i\beta\lrp{\mu_L+\mu_R}}, \\
	\lim_{\tau\to\infty}\frac{P(q)}{P(-q)} &= e^{q\beta\lrp{\mu_L+\mu_R}}.\label{pairXFT}
\end{align}

	We notice that the XFT~(\ref{pairXFT}) is different from the conventional XFT~(\ref{XFTtr}). Usually, the XFT is relevant to the difference of the chemical potentials of the two reservoirs~$\mu_L-\mu_R$. But in Eq.~(\ref{pairXFT}), it's relevant to the sum of the two chemical potential~$\mu_L+\mu_R$. One consequence of Eq.~(\ref{pairXFT}) is that for a symmetric bias $\mu_L=-\mu_R>0$, the average current $\avg{q}/\tau$ is zero. Eq.~(\ref{pairXFT}) can be understood intuitively by considering the entropy production. {\color{black}According to thermodynamics, the transport of particles is accompanied by the conversion of chemical energy to heat, and thus accompanied by the generation of entropy. When an electron is transported from the left~(right) reservoir to the grounded superconductor, the entropy production is $\Delta \Omega_L=\beta\mu_L$~$(\Delta \Omega_R = \beta\mu_R)$. Thus the total entropy production for a CAR process is 
\begin{align}
	\Delta \Omega = \Delta \Omega_L + \Delta \Omega_R= \beta(\mu_L+\mu_R).
\end{align}}
From Eq.~(\ref{entropy}), we recover Eq.~(\ref{pairXFT}). It's worth mentioning that this form of the XFT~(\ref{pairXFT}) is commonly found in the multiple terminal systems~\cite{Krause2011prb, Ganeshan2011prb, Cuetara2011prb}. 

The parity dependence of the transport can be understood intuitively as follows: as an electron jumps from the left reservoir to the first site of the Kitaev chain, it can form a Cooper pair with one electron on the second site. Then a hole is left on the second site and will pair with the hole on the third site. The pair forming process continues until the last site. It's easy to see that if there are odd-number sites in the Kitaev chain, the last site will host an electron and can jump to the right reservoir while a chain with even-number sites will transport a hole to the right reservoir. Thus, we conclude that in the paring case, a system with even parity gives rise to the CAR and the odd parity leads to normal transport.

We also calculate the long-time statistics of the current for $10$ (see Fig.~\ref{sc1011u0d1T0}(a),~\ref{sc1011u0d1T0}(b))  and $11$ (see Fig.~\ref{sc1011u0d1T0}(c), ~\ref{sc1011u0d1T0}(d)) sites in the paring case under the symmetric bias~($\mu_L=-\mu_R$) to illustrate the even-odd difference. As clearly demonstrated in Fig.~\ref{sc1011u0d1T0}(a), the rate function is symmetric with respect to $q/\tau=0$. Fig.~\ref{sc1011u0d1T0}(b) confirms the XFT~(\ref{pairXFT}) in the case of symmetric bias. Meanwhile, for 11 sites, Fig.~\ref{sc1011u0d1T0}(c) shows the net current $\avg{q}/\tau>0$ and the rate function exhibits asymmetry. The XFT for 11 sites is the conventional XFT~(\ref{XFTtr}), see Fig.~\ref{sc1011u0d1T0}(d).

\subsection{Majorana case}
Now we consider the exactly solvable case~($\mu=0, \Delta=\eta$), Majorana case, which has been extensively studied in the literature \cite{Roy2012prb, Doornenbal2015prb, Bhat2020prb}. In this case, MZMs will appear at the ends of the Kitaev chain. We find the CF is given by
\begin{align}
	Z_3(\xi, \omega) = \bbA_{3}(\omega)+\bbC_{3}(\omega)\lrp{\barn_{1e} n_{1h} e^{-2i\xi} +n_{1e}\barn_{1h} e^{2i\xi}}.\label{mzmCF}
\end{align}
{\color{black}It's worth mentioning that the explicit expression of $\bbA_3(\omega)$ and $\bbC_3(\omega)$~(see Appendix~\ref{coefficient}) are independent of the site number in this exactly solvable case. The Kitaev chain behaves like a three level system~(see Appendix~\ref{three-level}).} 

From Eq.~(\ref{mzmCF}), we can see that the transport between the left reservoir and the chain is independent of the transport between the chain and the right reservoir. {\color{black} In fact, if one also adds a counting field corresponding to the measured particle number of the right reservoir, one would see that the CF will factorize into two parts, corresponding to the left and the right part respectively. So we conclude that there is no correlation between the particle currents at the two interfaces.} 

Let's consider the zero temperature and the symmetric bias. The CGF becomes
\begin{align}
	\mcF_3(\xi) = \int_{-\mu}^{\mu}\frac{d\omega}{4\pi}\ln Z_3(\xi,\omega). 
\end{align}
And accordingly, the zero-bias conductance is 
\begin{align}
	G_L &= \frac{\partial \avg{I}}{\partial \mu}\bigg|_{\mu\to 0} = \frac{-i}{2\pi}\frac{\partial}{\partial \xi}\lrb{\ln Z_3(\xi,0)}\bigg|_{\xi\to 0} \nonumpar
	&= \frac{1}{\pi} = \frac{2e^2}{h},
\end{align}
where in the last step we return to the SI units and recovers the famous quantized zero-bias conductance of the MZMs. 

Eq.~(\ref{mzmCF}) indicates the fundamental process is a local Andreev reflection~(LAR), in which an electron is  injected into the chain and a hole is reflected out, and a Cooper pair is formed in the chain. The XFT for this process is 
\begin{align}
	\lim_{\tau\to\infty}\frac{P(q)}{P(-q)} = e^{q\beta \mu_L}.  \label{tscxft}
\end{align}

Fig.~\ref{tscu0d1}(b) demonstrates the validity of Eq.~(\ref{tscxft}). Fig.~\ref{tscu0d1}(a) shows that only even-number particles can be transported from the left reservoir to the Kitaev chain. {\color{black} This is expected since the only transport process happened in the Majorana case is the LAR process, which exchanges two electrons effectively. It can also be seen from the CF~(\ref{mzmCF}) which has an enhanced periodicity 
\begin{align}
	Z_3(\xi, \omega) = Z_3(\xi+\pi, \omega),
\end{align}
compared to the conventional $2\pi$-periodicity in the CF~(\ref{trcf}, \ref{pair3cf}, \ref{pair4cf}). From the definition of $Z(\xi)=\sum_q \exp(iq\xi)P(q)$ and the normalization of the probability $\sum_q P(q)=1$, we have 
\begin{align}
	Z_3(0) &= \sum_q P(q) = \sum_{q\in \rmEven}P(q)+\sum_{q\in\rmOdd}P(q)=1, \\
	Z_3(\pi) &= \sum_q e^{iq\pi}P(q) =\sum_{q\in \rmEven}P(q)-\sum_{q\in\rmOdd}P(q) \nonumpar
	&= Z_3(0) = 1.
\end{align} 
From the above two equations, we find 
\begin{align}
	\sum_{q\in\rmEven} P(q)=1, \quad \sum_{q\in\rmOdd} P(q)=0
\end{align}
which implies that it's impossible to transport odd-number particles. 
} 

\subsection{General case}
From the above discussion, we see that there exists three distinct current components in the transport process, the normal transport, the CAR and the LAR. Each current component has a corresponding XFT. When we consider a generic situation, i.e., $\mu\neq 0, \eta\neq 0, \Delta\neq 0$, we expect three current components coexist in the transport process. We take $\mu=\eta=\Delta=1$ as an example. The $3$-site CF is 
\begin{widetext}
\begin{align}
	Z_3(\xi,\omega) 
	= \bbA_4(\omega) 
	&+ \lrb{\lrp{\bbC_{41}\barn_{1e}n_{2e}+\bbC_{42} n_{1h}\barn_{2h}}+\lrp{\bbC_{43}\barn_{1e}n_{2h}+\bbC_{44} n_{1h}\barn_{2e}}}e^{-i\xi}+ \bbB_4(\omega)\lrp{n_{1e}\barn_{1h}}e^{2i\xi} \nonumpar
	&+ \lrb{\lrp{\bbC_{41} n_{1e}\barn_{2e}+\bbC_{42} \barn_{1h}n_{2h}}+\lrp{\bbC_{43} n_{1e}\barn_{2h}+\bbC_{44} \barn_{1h}n_{2e}}}e^{i\xi}+ \bbB_4(\omega)\lrp{n_{1h}\barn_{1e}}e^{-2i\xi}, \label{gencf}
	\end{align}
\end{widetext}
where the coefficients $\bbA_4, \lrc{\bbC_{4j}}_{j=1}^4$ and $\bbB_4$ are given in Appendix $\ref{coefficient}$.
We see that all the three current components present in the transport process, thus rendering no simple XFT for the total current from a single reservoir. Since the TPM on the left reservoir only gives us the total current from the left reservoir, we need TPMs on the other reservoir and also the superconductor to gain additional information. It's possible to carry out the TPMs on three terminals simultaneously~(since all the particle number operators commute). Denote the outcomes as $q_L,q_R$ and $q_S$. The three outcomes are not independent. Charge conservation demands $q_L+q_R+q_S=0$ since charge concentration can't occur in the chain in the steady state. The conventional steady-state XFT of two independent outcomes $q_L, q_R$ is given by ~\cite{Campisi2011rmp, Andrieux2009njp, Gaspard2013book} 
\begin{align}
	\lim_{\tau\to \infty}\frac{P(q_L,q_R)}{P(-q_L,-q_R)} = e^{\beta\lrp{q_L\mu_L+q_R\mu_R}}\label{fullxft}
\end{align}
which concerns the net change of electron number in {\color{black}each} reservoir. But we can also consider how many electrons are transported in different current components, which corresponds to charge transfer in the normal transport $q_n$, the CAR process $q_c$, the left LAR process $q_l$, and the right LAR process $q_r$, respectively. These components are related to $q_l$ and $q_r$ by the following equations
\begin{align}
	q_n + q_c + 2 q_l &= q_L,\nonumpar
	-q_n + q_c + 2 q_r &= q_R.
\end{align}
Unfortunately, we have four independent variables in the left-hand side but only two independent equations. It implies that  we can't know the exact particle number transported in different current components except some special cases. For example, if there are no LARs, then both $q_l$ and $q_r$ are zero, and we have 
\begin{align}
	q_c = \frac{q_L+q_R}{2}, \quad q_n = \frac{q_L-q_R}{2}. 
\end{align}
The XFT (\ref{fullxft}) can be rewritten as
\begin{align}
	\lim_{\tau\to \infty}\frac{P(q_c,q_n)}{P(-q_c,-q_n)} = e^{\beta q_c\lrp{\mu_L+\mu_R}}e^{\beta q_n\lrp{\mu_L-\mu_R}}.
\end{align}

\begin{figure}
\includegraphics[scale=0.65]{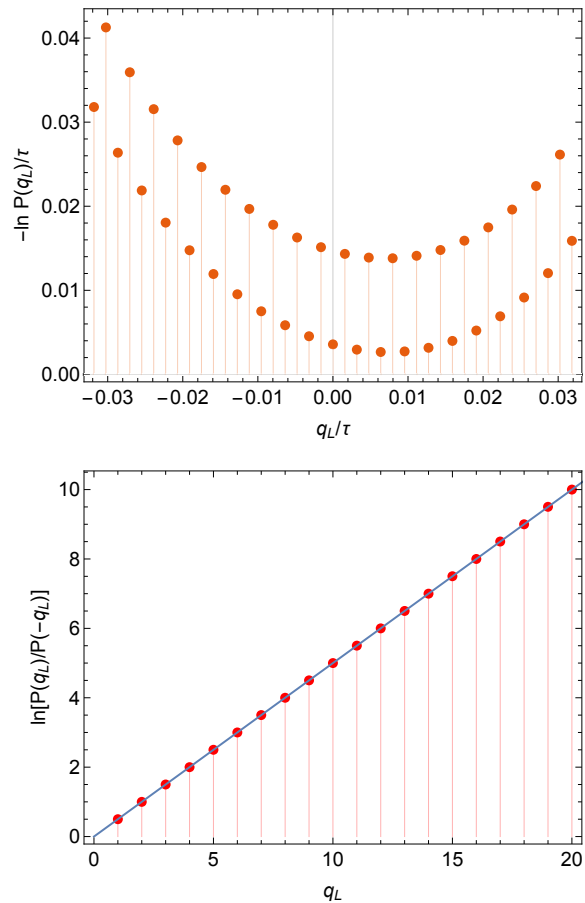}
\caption{The rate function and XFT for 10 sites in the general case. The parameters are: $\mu=\Delta=\eta=1$, $\Gamma_L=\Gamma_R=0.3$, $\mu_L=-\mu_R=0.05$, $\beta=10$, $\Delta\omega \equiv 2\pi/\tau=0.01$. (a) The rate function $-\ln P(q_L)/\tau$; (b) The ratio $\ln P(q_L)/\ln P(-q_L)$. The solid line in (b) is around $0.500073q$, showing a deviation from the above three XFTs~(\ref{XFTtr},~\ref{pairXFT},~\ref{tscxft}).}
\label{gen10u1d1}
\end{figure}

We numerically calculate the rate function for 10 sites. The rate function for even and odd particle transport exhibits distinct features and indicates the three transport current components coexist and the dominant one is the LAR process. From Fig.~\ref{gen10u1d1}(b), we note that the ratio $\ln P(q_L)/\ln P(-q_L)\sim q_L\beta\mu_L$, showing a small deviation from the three exact XFTs~(\ref{XFTtr},~\ref{pairXFT},~\ref{tscxft}). 


\section{Discussion and conclusion}
Previously, quantum transport through a Kitaev chain in the continuum limit and$/$or zero temperature has been extensively studied. But the focus is usually put on the mean current and the shot noise, which are experimentally accessible with the current experimental techniques. Nevertheless, higher-order moments are usually ignored in previous studies. Also, the transport through a finite Kitaev chain at a finite temperature, {\color{black} and the influence of the $\mcU(1)$ symmetry breaking term in the Kitaev chain on the forms of the XFTs, has not been fully explored. In our current study, we elaborate the influence of the $\mcU(1)$ symmetry breaking} on the XFTs of currents through a Kitaev chain. Thus, our work establishes a connection between studies of the transport in a Kitaev chain and studies of the XFTs. Hopefully, our results can be experimentally verified in future experiments.

{\color{black}In this article, we focus on the FTs of particle transport. And the transport of particles is always accompanied by heat flux. Hence, we may also consider the FTs regarding heat flux. It turns out that various forms of FTs of particle transport correspond to a unique heat FT, which is the same as Eq.~(1) except replacing $\Delta \Omega$ with $T\Delta\Omega$.}

In summary, we calculate the full counting statistics of the electron transport through a short Kitaev chain. By utilizing the Keldysh formalism, we obtain the analytical expression of the generating function of currents. In principle, the analytical expression of the generating function is valid for arbitrary number of sites. But in practice, the calculation becomes more and more tedious when the site number increases. Also we would like to point out that from these analytical results, we can straightforwardly demonstrate different current components satisfying various XFTs, which is consistent with the framework of entropy production. The four cases considered in our model manifest the influence of the {\color{black} $\mcU(1)$ symmetry breaking on} the properties of the quantum transport and the XFTs. {\color{black}In the paring case and Majorana case, the paring terms in the Hamiltonian, which breaks the $\mcU(1)$ symmetry, lead to two kinds of Andreev reflections, resulting in two unconventional XFTs~(Eq.~(\ref{pairXFT},~\ref{tscxft})).} It's worth mentioning that analytical results can be obtained in our study because the Hamiltonian of the Kitaev chain is in the quadratic form. For systems beyond the quadratic Hamiltonian, one has to refer to the perturbation expansion or numerical methods. In our future study, we plan to extend our investigations to Hamiltonian with non-quadratic terms, which is ubiquitous in physics and may lead to many interesting phenomena. 

\begin{acknowledgments}
H. T. Quan acknowledges support from the National Science Foundation of China under grants 11775001, 11534002, and 11825001.
\end{acknowledgments}

\appendix   

\section{The coefficients of the CF in different cases}\label{coefficient}

In this Appendix, we give the expressions of the coefficients of the CF~{\color{black}(\ref{trcf}, \ref{pair3cf}, \ref{pair4cf}, \ref{mzmCF}, \ref{gencf}).}
The coefficients $\bbA_{1\pm}$ and $\bbC_{1\pm}$ for the trivial case in Eq.~(\ref{trcf}) are given by
\begin{widetext}
\begin{align}
	\bbA_{1+} = &\frac{1}{\bbD_{1+}}\big[\Gamma _L^2 \left(\eta ^4-2 \eta ^2 (\mu +\omega )^2+(\mu +\omega )^4+(\mu +\omega )^2 \Gamma _R^2\right)+\Gamma _R^2 \left(\eta ^2-(\mu +\omega )^2\right)^2 \nonumpar
	&+2 {\color{black}(1-2n_{1e}\barn_{2e}-2\barn_{1e}n_{2e})} \eta ^4 \Gamma _L \Gamma _R+(\mu +\omega )^2 \left((\mu +\omega )^2-2 \eta ^2\right)^2\big],\\
	\bbA_{1-} = &\frac{1}{\bbD_{1-}}\big[ \Gamma _L^2 \left(\eta ^4-2 \eta ^2 (\mu -\omega )^2+(\mu -\omega )^4+(\mu -\omega )^2 \Gamma _R^2\right) +\Gamma _R^2 \left(\eta ^2-(\mu -\omega )^2\right)^2 \nonumpar
	&+ 2 {\color{black}(1-2n_{1h}\barn_{2h}-2\barn_{1h}n_{2h})} \eta ^4 \Gamma _L \Gamma _R+(\mu -\omega )^2 \left((\mu -\omega )^2-2 \eta ^2\right)^2\big], \\
	\bbC_{1\pm} = & \frac{4 \eta ^4 \Gamma _L \Gamma _R}{\bbD_{1\pm}}, \\
	\bbD_{1\pm} =& \Gamma _L^2 \left(\eta ^4-2 \eta ^2 (\mu \pm\omega )^2+(\mu \pm\omega )^4+(\mu \pm\omega )^2 \Gamma _R^2\right)+\Gamma _R^2 \left(\eta ^2-(\mu \pm\omega )^2\right)^2 +2 \eta ^4 \Gamma _L \Gamma _R\nonumpar
	 &+ (\mu \pm \omega )^2 \left((\mu \pm \omega )^2-2 \eta ^2\right)^2.
\end{align}	
\end{widetext}


The coefficients $\bbA_{2\pm}, \bbA'_{2\pm}, \bbC_{2\pm}$ and $\bbC'_{2\pm}$ for the paring case in Eq.~(\ref{pair3cf}) and Eq.~(\ref{pair4cf}) are given by
\begin{widetext}
\begin{align}
	\bbA_{2+}(\omega) = & \frac{1}{\bbD_{2+}} \big[(\mu +\omega )^2 \left(2 \Delta ^2+\mu ^2-\omega ^2\right)^2+2 \Delta ^4 {\color{black} (1-2n_{1h}\barn_{2h}-2\barn_{1h}n_{2h})}  \Gamma _L \Gamma _R+\Gamma _R^2 \left(\Delta ^2+\mu ^2-\omega ^2\right)^2 \nonumpar
	&+\Gamma _L^2 \left(\left(\Delta ^2+\mu ^2-\omega ^2\right)^2+(\mu -\omega )^2 \Gamma _R^2\right), \\
	\bbA_{2-}(\omega) = &\frac{1}{\bbD_{2-}}\big[(\mu -\omega )^2 \left(2 \Delta ^2+\mu ^2-\omega ^2\right)^2+2 \Delta ^4 {\color{black} (1-2n_{1e}\barn_{2e}-2\barn_{1e}n_{2e})} \Gamma _L \Gamma _R+\Gamma _R^2 \left(\Delta ^2+\mu ^2-\omega ^2\right)^2 \nonumpar
	&+\Gamma _L^2 \left(\left(\Delta ^2+\mu ^2-\omega ^2\right)^2+(\mu +\omega )^2 \Gamma _R^2\right) \big], \\	
	\bbA_{2+}'(\omega) = &\frac{1}{\bbD_{2+}'}\big[\Delta ^8+6 \Delta ^6 \left(\mu ^2-\omega ^2\right)+2 \Delta ^6 {\color{black}(1-2n_{1h}\barn_{2e}-2\barn_{1h}n_{2e})} \Gamma _L \Gamma _R+(\mu -\omega )^2 (\mu +\omega )^4 \left((\mu -\omega )^2+\Gamma _R^2\right) \nonumpar
	 &+ \Gamma _L^2 \left((\mu -\omega )^2 \left(2 \Delta ^2+\mu ^2-\omega ^2\right)^2+\Gamma _R^2 \left(\Delta ^2+\mu ^2-\omega ^2\right)^2\right)+\Delta ^4 (\mu +\omega )^2 \left(11 (\mu -\omega )^2+4 \Gamma _R^2\right) \nonumpar
	 &+2 \Delta ^2 (\mu -\omega ) (\mu +\omega )^3 \left(3 (\mu -\omega )^2+2 \Gamma _R^2\right)\big], \\	
	\bbA_{2-}'(\omega) = & \frac{	1}{\bbD_{2-}'}\big[\Delta ^8+6 \Delta ^6 \left(\mu ^2-\omega ^2\right)+2 \Delta ^6 {\color{black}(1-2n_{1e}\barn_{2h}-2\barn_{1e}n_{2h})} \Gamma _L \Gamma _R+(\mu -\omega )^4 (\mu +\omega )^2 \left((\mu +\omega )^2+\Gamma _R^2\right) \nonumpar
	&+\Gamma _L^2 \left((\mu +\omega )^2 \left(2 \Delta ^2+\mu ^2-\omega ^2\right)^2+\Gamma _R^2 \left(\Delta ^2+\mu ^2-\omega ^2\right)^2\right)+\Delta ^4 (\mu -\omega )^2 \left(11 (\mu +\omega )^2+4 \Gamma _R^2\right) \nonumpar
	&+ 2 \Delta ^2 (\mu -\omega )^3 (\mu +\omega ) \left(3 (\mu +\omega )^2+2 \Gamma _R^2\right)\big], \\
	\bbC_{2\pm} (\omega) = & \frac{4 \Delta ^4 \Gamma_L \Gamma_R}{\bbD_{2\pm}}, \quad	
	\bbC'_{2\pm} (\omega) = \frac{4 \Delta ^6  \Gamma_L \Gamma_R}{\bbD_{2\pm}'},
\end{align}
\end{widetext}
where $\bbD_{2\pm}$ and $\bbD_{2\pm}'$ are given by 
\begin{widetext}
\begin{align}
	\bbD_{2\pm} =& (\mu \pm \omega )^2 \left(2 \Delta ^2+\mu ^2-\omega ^2\right)^2+2 \Delta ^4 \Gamma _L \Gamma _R+\Gamma _L^2 \left(\left(\Delta ^2+\mu ^2-\omega ^2\right)^2+(\mu \mp\omega )^2 \Gamma _R^2\right)+\Gamma _R^2 \left(\Delta ^2+\mu ^2-\omega ^2\right)^2, \\
	\bbD_{2\pm}' =& (\mu \mp\omega )^2 (\mu \pm \omega )^4 \left((\mu \mp \omega )^2+\Gamma _R^2\right)+2 \Delta ^6 \Gamma _L \Gamma _R+\Gamma _L^2 \left((\mu \mp \omega )^2 \left(2 \Delta ^2+\mu ^2-\omega ^2\right)^2+\Gamma _R^2 \left(\Delta ^2+\mu ^2-\omega ^2\right)^2\right) \nonumpar
	&+\Delta ^4 (\mu \pm \omega )^2 \left(11 (\mu \mp \omega )^2+4 \Gamma _R^2\right)+2 \Delta ^2 (\mu \mp \omega ) (\mu \pm \omega )^3 \left(3 (\mu \mp\omega )^2+2 \Gamma _R^2\right)+\Delta ^8+6 \Delta ^6 \left(\mu ^2-\omega ^2\right).
\end{align}	
\end{widetext}

{\color{black}The coefficients $\bbA_3$, $\bbC_3$ for the Majorana case in Eq.~(\ref{mzmCF}) are given by 
\begin{widetext}
\begin{align}
	\bbA_3(\omega) &= \frac{\Gamma_L^4 \omega ^2+2 \Gamma_L^2 \left(8{\color{black}(1-n_{1e}\barn_{1h}-\barn_{1e}n_{1h})}+\omega ^4-4 \omega ^2\right)+\omega ^2 \left(\omega ^2-4\right)^2}{\Gamma_L^4 \omega ^2+2 \Gamma_L^2 \left(\omega ^4-4 \omega ^2+8\right)+\omega ^2 \left(\omega ^2-4\right)^2}, \label{a3}\\
	\bbC_3(\omega) &= \frac{16 \Gamma_L^2}{\Gamma_L^4 \omega ^2+2 \Gamma_L^2 \left(\omega ^4-4 \omega ^2+8\right)+\omega ^2 \left(\omega ^2-4\right)^2}.\label{c3}
\end{align}
\end{widetext}
}
The coefficients $\bbA_{4}$, $\lrc{\bbC_{4j}}_{j=1}^4$ and $\bbB_4$ for the general case in Eq.~($\ref{gencf}$) are given by{\color{black}
\begin{widetext}
\begin{align}
	\bbA_4 (\omega)&= \frac{1}{\bbD_{4}}\big[\bbD_4 - \lrp{\bbC_{41}\barn_{1e}n_{2e}+\bbC_{42} n_{1h}\barn_{2h}}-\lrp{\bbC_{43}\barn_{1e}n_{2h}+\bbC_{44} n_{1h}\barn_{2e}}-\bbB_4 n_{1e}\barn_{1h} \nonumpar
	& \quad \quad ~~~- \lrp{\bbC_{41} n_{1e}\barn_{2e}+\bbC_{42} \barn_{1h}n_{2h}}-\lrp{\bbC_{43} n_{1e}\barn_{2h}+\bbC_{44} \barn_{1h}n_{2e}}-\bbB_4 n_{1h}\barn_{1h}\big], \\
	\bbC_{41}(\omega) &= \frac{1}{ \bbD_{4}}\big[16 \Gamma_L \Gamma_R \lrp{\Gamma_L^2+(\omega +1)^2} \lrp{\Gamma_R^2+\lrp{1+\omega}^2}\big], \\
	\bbC_{42}(\omega) &= \frac{1}{ \bbD_{4}}\big[16 \Gamma_L \Gamma_R\lrp{\Gamma_L^2+(\omega -1)^2} \lrp{\Gamma_R^2+\lrp{1-\omega}^2}\big], \\
	\bbC_{43}(\omega) &= \frac{1}{ \bbD_{4}}\big[16 \Gamma_L \Gamma_R \lrp{\Gamma_L^2+(\omega +1)^2} \lrp{\Gamma_R^2+\lrp{1-\omega}^2}\big], \\
	\bbC_{44}(\omega) &= \frac{1}{ \bbD_{4}}\big[16 \Gamma_L \Gamma_R\lrp{\Gamma_L^2+(\omega -1)^2} \lrp{\Gamma_R^2+\lrp{1+\omega}^2}\big], \\
	\bbB_4(\omega) &= \frac{16 \Gamma_L^2 \lrb{\omega ^2 \Gamma_R^4+2 \left(\omega ^4-3 \omega ^2+8\right) \Gamma_R^2+\omega ^2 \left(\omega ^2-5\right)^2}}{\bbD_{4}},
\end{align}
\end{widetext}
}
where $\bbD_{4}$ is given by
\begin{widetext}
{\color{black}
\begin{align}
	\bbD_4(\omega) &= \left(\omega ^6-11 \omega ^4+27 \omega ^2-1\right)^2 +2 \left(\omega ^{10}-15 \omega ^8+78 \omega ^6-178 \omega ^4+209 \omega ^2+1\right) \left(\Gamma _L^2+\Gamma _R^2\right) \nonumpar
	&+\Gamma _R^4 \left((\omega -1)^2 \Gamma _L^2+((\omega -2) \omega -1)^2\right) \left((\omega +1)^2 \Gamma _L^2+(\omega  (\omega +2)-1)^2\right) \nonumpar
	&+\left(\omega ^4-6 \omega ^2+1\right)^2 \Gamma _L^4+32 \Gamma _L \Gamma _R \left(\Gamma _L^2+\omega ^2+1\right) \left(\Gamma _R^2+\omega ^2+1\right) \nonumpar
	&+2 \Gamma _R^2 \left(\left(\omega ^6-5 \omega ^4+11 \omega ^2+1\right) \Gamma _L^4+2 \left(\omega ^8-8 \omega ^6+30 \omega ^4-48 \omega ^2+65\right) \Gamma _L^2\right)
\end{align}
}
\end{widetext}

\section{The exactly solvable case}\label{three-level}
In the exactly solvable case $\mu=0, \eta=\Delta$, the Kitaev chain reduces to a free fermion model in the Majorana representation
\begin{align}
	\hatH_K=i\eta\sum_{j=1}^{N-1}\hat{\gamma}_{2j}\hat{\gamma}_{2j+1} = 2\eta\sum_{j=1}^{N-1}\hat{d}^{\dagger}_j\hat{d}_j+ 0 \cod_0\aod_0. \label{mzmH}
\end{align}
The fermionic operator $\hat{c}_j, \hat{d}_j$ is given by 
\begin{align}
	\hat{c}_j &= \frac{(\hat\gamma_{2j-1}+i\hat\gam_{2j})}{2}, 
	\hat{d}_j = \frac{\hat{\gamma}_{2j}+i\hat\gamma_{2j+1}}{2}\\
\hat{d}_0 &= \frac{\hat{\gamma}_{1}+i\hat\gamma_{2N}}{2}.
\end{align}
where the Majorana operator $\hat{\gamma}_j$ satisfies the anti-commutation relation $\{\hat{\gamma}_i,\hat{\gamma}_j\}=2\delta_{ij}$.
The coupling between the chain and the reservoir in the new representation is 
\begin{align}
	\coc_{Lj} \aoc_1 +h.c. &= \coc_{Lj} \frac{\cod_1+\aod_1+\aod_0-\cod_0}{2} +h.c., \nonumpar
	\coc_{Rj} \aoc_N + h.c. &= \coc_{Rj} \frac{\cod_0+\aod_0+\aod_{N-1}-\cod_{N-1}}{2} + h.c. . \label{mzmcoup}
\end{align} 
From Eq.~(\ref{mzmcoup}), we see the degrees of freedom of the reservoir only couple to three modes of the chain $\cod_1, \aod_1, \aod_{N-1},\cod_{N-1}, \cod_0$ and $\aod_0$. Since Eq.~(\ref{mzmH}) describes a free fermion Hamiltonian, we conclude that the Kitaev chain is an effective three level system in this exactly solvable case. 



\bibliography{ref}

\end{document}